# A Forecast Based Load Management Approach For Commercial Buildings – Comparing LSTM And Standardized Load Profile Techniques

Thomas Steens[a], Jan-Simon Telle[a], Benedikt Hanke[a],

Karsten von Maydell[a], Carsten Agert[a], Gian-Luca di Modica[b], Bernd Engel[b] and Matthias Grottke[c]

[a] DLR Institute of Networked Energy Systems, Carl-von-Ossietzky-Str. 15, 26129 Oldenburg, Germany

[b]Technische Universität Braunschweig Institut für Hochspannungstechnik und Elektrische Energieanlagen - elenia, Schleinitzstraße 23, 38106, Braunschweig, Germany

[c]Hammer Real GmbH, Sylvensteinstr. 2, 81369 Munich, Germany

**ABSTRACT**: Load-forecasting problems have already been widely addressed with different approaches, granularities and objectives. Recent studies focus not only on deep learning methods but also on forecasting loads on single building level. This study aims to research problems and possibilities arising by using different load forecasting techniques to manage loads. For that the behaviour of two neural networks, Long Short-Term Memory and Feed Forward Neural Network and two statistical methods, standardized load profiles and personalized standardized load profiles are analysed and assessed by using a sliding-window forecast approach. The results show that machine learning algorithms have the benefit of being able to adapt to new patterns, whereas the personalized standardized load profile performs similar to the tested deep learning algorithms on the metrics. As a case study for evaluating the support of load-forecasting for applications in energy management systems, the integration of charging stations into an existing building is simulated by using load forecasts to schedule the charging procedures. It shows that such a system can lead to significantly lower load peaks, exceeding a defined grid limit, and to a lower number of overloads compared to uncontrolled charging.

# 1. INTRODUCTION

The on-going energy system transformation process, intended to reduce $CO_2$ emissions and meet the EU's long-term goal of being climate-neutral by 2050, is shifting the dependence of energy production from fossil fuels to renewable energy sources such as wind and solar energy. This also leads to the integration of the three main energy sectors of electricity, heat and transport into buildings. For this reason, new challenges arise for electricity grids. As far as energy production is concerned, the increasing share of renewable energy sources and the resulting dependency on them [1] introduces uncertainties into power generation [2].

A further problem for the electricity grids arises from the electrification of the transport sector through increasing overall electricity consumption and with charging times overlapping with periods of high peak loads [3]. It is expected that the charging of electric cars will have an effect on the general network stability in Germany with a share of 10-20%. However, due to the expected peak loads, a load management system can avoid overloading the grid connection point or the upstream transformer [4]. As an example, the integration of charging infrastructure for battery-electric vehicles (BEVs) into the existing building stock will create high demand for load management in order to avoid infrastructure extension.

A solution for increasing grid stability when integrating charging facilities for BEVs into existing buildings can be a load forecast-based load management system. With such a system, charging processes can be scheduled and e.g., shifted to times when building load balances fit power restrictions. From the view of an electric utility such a system also enables the shifting of loads to beneficial times when the consumer participates in demand response tariffs [5].

Load-forecasting has been researched for many years with model- and data-driven approaches. One model-driven approach is the use of statistical *standardized load profiles* (SLP), which were derived using the measured loads of different buildings in 15 minute intervals in 1999 [6]. Due to the now emerging smart grids and planned smart meter rollout at the building level, load forecasts with a significantly higher resolution (< 15 minutes) can be provided. In a study, the accuracy of load forecasting for regions was significantly improved with the use of *personalized standardized load profiles* (PSLPs) because of their incorporation of on-site measured data [7].

Many forecasting techniques have already been tested to forecast loads on different system levels. Traditional load-forecasting techniques based on regression and time series analyses such as autoregressive integrated moving average (ARIMA) in [8,9] multiple regression in [10,11] or support vector regression [12,13], have already been applied. With the ever-increasing computing power Deep Learning (DL) methods are being tested and used, as they have proved effective at solving different problems in text and language processing, as well as image recognition. Algorithms such as feed-forward neural networks (FFNNs) [14,15] and Long Short-Term Memory (LSTM) [16,17] are applied because of their ability to adapt to nonlinear problems and the possibility of computing results by means of big datasets.

Lindberg et Al. [18] predicted aggregated energy consumption of different non-residential buildings by using regression models and data of outdoor temperature, time of day and type of day. Bento et Al. [19] used an LSTM network via an improved Bat Algorithm to perform weekly regional system loads forecasts. Kong et Al. [20] showed that by going from substation level to single building level the energy consumption becomes volatile. This lowers the forecasting performance for residential buildings as the proposed LSTM struggled to perform well on the test data. In contrast to residential building load profiles large commercial buildings have a more stable usage pattern, because one action within the building leads to minor changes in the load profiles However the smaller the building the more a single action can cause a higher effect on the load patterns. [21] Load forecasting for commercial buildings has been compared to predicting residential loads with recurrent neural networks by Rahman et. Al.[22]. It showed that predicting loads on single residential building level leads to high forecasting errors compared to commercial buildings or aggregated residential loads. This is due to load patterns becoming more distinct. Thokala et Al. [23] compared linear regression methods to SVR and Non-linear autoregressive neural network with exogenous output for commercial building load forecasting and stated out that both are performing better than the linear regression. Nichiforov et Al. . [24] showed that for large commercial buildings recurrent neural networks with a layer of LSTM can achieve accurate forecasts.

This paper presents an approach to assess the behaviour of load forecasting techniques and investigates the capabilities of using the recently more and more used deep learning techniques for load management applications. For that basic LSTM and FFNN models are optimized and trained in a sliding window forecasting framework as e.g. used and proposed by Bedi et. Al. [25] The forecasting results are compared to the in Germany still used statistical methods (SLP and PSLP) to quantify problems and advantages of the different methods in their forecasting behaviour and accuracy. Widely approved and important features were used to derive the models for the neural networks. Weather parameters such as temperature, humidity, wind speed, etc., were shown to be important and widely used variables [26,27]. The ambient air temperature for example, has a high correlation with the load measurements of a building, while humidity also has a direct impact on people's energy consumption behaviour in buildings, such as the need for increased cooling [28]. Calendar effects such as indicators for the day or time are also used, but tend to have a low impact on the outcome [29]. Also because of the arising possibility of using data with higher resolution opened by smart metering systems the granularity of the data is lowered to a 5 minute resolution scale. Combining the sliding window and the decrease of granularity allows the exploration of load forecasting based load management systems in greater detail.

The forecast accuracy was measured by different commonly used absolute metrics, such as *Mean Absolute Error* (MAE) and *Root Mean Squared Error* (RMSE), and normalized metrics, e.g. *Mean Absolute Percentage Error* (MAPE) (used for example by Hossen et Al. [30] and Fen et Al. [31]). Also in this study the more recently introduced *Mean Absolute Scaled Error* (MASE); described by Hyndman et Al. [32] is used. Only the simple SLP falls out of rank, while the PSLP and FFNN perform in a similar way.

The necessity of forecast-based load management was discussed in a case study that is schematically shown in. Therefore, the integration of a varying number of BEVs with different charging behaviours and a fixed power limit in the building were simulated and evaluated. It showed the possibility of integrating a larger amount of charging infrastructure into existing commercial buildings by applying forecast-based charging strategies.

The key contributions of this paper are:

- Using a sliding window approach to continuously forecast loads with a granularity of 5 min
- Develop and optimize deep neural network models for 24-hours ahead load prediction with intraday updates in the simulation framework.
- Quantify the performance of these models and compare them to standardized load profile techniques concerning their forecasting capabilities and behaviour towards load management applications.
- Case study of integration of electric vehicles into an existing commercial building.
- Load management: Comparison between unscheduled and forecast based BEV charging.

Section 2 features the description of datasets of measured commercial building loads, the data exploration and a description of the features for the neural networks. In section 3 the sliding window framework for load forecasting is described whereas in section 4 the use-case of charging battery electric vehicles as dynamic consumers in the commercial buildings is introduced. In section 5 and 6 the results concerning load forecasting and the use case respectively are presented and discussed. A conclusion and suggestions for future works are summarized in the final section.

## 2. DATA DESCRIPTION

The algorithms used in this study are data-driven approaches. Therefore, it is important to evaluate and understand the data. The data is explored to assess the quality of the datasets used and different features are defined to be utilized in the ML process.

### 2.1. Dataset Description

In this study two datasets of measured loads from different commercial buildings in Germany are used (Table 1). The data is collected over a period of seven months (01/12/2018 to 30/06/2019) at a one second time resolution. One of the two datasets is used as the *main dataset* (MD), while the other is used as a *validation dataset* (VD) to validate the results. As the datasets do not include a complete year, the annual energy consumption had to be estimated. For this purpose, the average output in the measurement period was calculated and related to the entire year. For the optimization of the machine learning algorithms and the validation, the datasets were shortened to a period from 02/03/2019 to 03/04/2019. This period was chosen because of the absence of public vacations, making it an idealized dataset containing only working days and weekends.

Table 1. Characteristics of the main and validation datasets

| Dataset | MD | VD |
|---|---|---|
| **Time period** | 01/12/2018 - 30/06/2019 | 02/03/2019 - 03/04/2019 |
| **Time resolution (after reduction)** | 5 minutes | 5 minutes |
| **Missing, double and incorrect data points [%]** | 0.2 | 3 |
| **Average load [kW]** | 19.89 | 3.12 |
| **Maximum load [kW]** | 84.74 | 10.43 |
| **Approx. annual load [MWh/a]** | 174.24 | 27.33 |

The load patterns in MD are constant throughout the measured period (Figure 1, upper), with around 50 kW on average and time-inconsistent peaks of up to 70 kW during working hours and an otherwise measured base load of 3.5 kW with peaks to 7 kW in non-working hours.

In the second commercial building for the VD (Figure 1, lower), on weekdays the load pattern is nearly constant with a load profile of between 4 kW and 8 kW in the working hours and peaks that exceeds 10 kW. The weekends are inconsistent, with most having high and fluctuating loads, while on one weekend, the base load was measured.

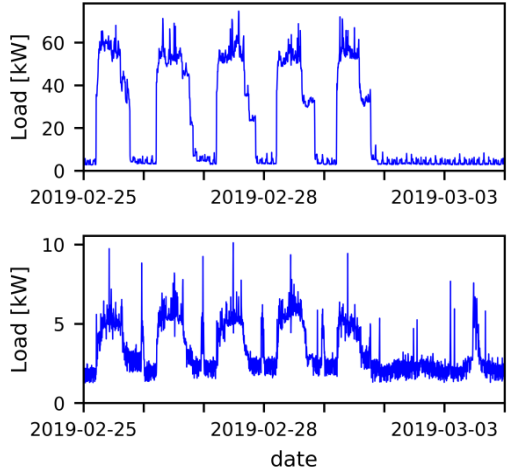

Figure 1. Example of the loads within commercial buildings (upper MD, lower VD)

In comparison, the load profiles of the two datasets considerably differ from each other. While loads are more constant in the MD, the loads in the VD building vary continually. The weekends are not as clear-cut for the VD building as for the MD building.

### 2.2. Data Exploration
Data-driven methods for predictions heavily rely on the quality of data provided. Therefore, both datasets must be checked for inconsistencies such as missing or duplicate values. In the raw datasets of the MD, measurements are missed between 05/04/2019 and 08/04/2019, and between 12/04/2019 and 14/04/2019. That renders a total of 3% of the data missing and 0.2 % in the VD.

### 2.3. Feature Description
Throughout the literature many different features were introduced and many have proven to be valuable for load forecasting as described in section 1. The features (Table 2) used in this study are divided into three groups:

1. Historical features
2. External features
3. Calendrical and timely features

For historical features, two characteristics are used: the measured load of the previous week and of previous day at exactly the same time as the observed data point.

As an external feature, the temperature is used, which has been shown to have a large impact on the load and a good correlation with it [33]. As there is no data measured on site, data from the "Deutscher Wetterdienst" (DWD), or German Meteorological Service are used [34].

In the group of calendrical and timely features, six different factors are defined: The first is the day of the week, initiated by the numbers 1 to 7. The difference between weekdays and weekends, but also between weekdays and holidays, is represented by 0 and 1 and included as additional features. The last two characteristics used are the sine and cosine function, which have a complete cycle once a day or week, respectively, and introduce periodicity into the model.

Table 2. Selected features for load forecasts done by ML-algorithms

| Feature Group | Feature Name | Feature Description |
|---|---|---|
| Historical Features | Measurements of week before [W] | Measured loads at the exact same time the week before |
| | Measurements of day before [W] | Measured loads at the exact same time 24 hours before |
| External features | Ambient Air Temperature [°C] | Actual Ambient Air Temperature |
| Calendrical and timely features | Day indicator [1-7] | Day of the Week by number |
| | Weekend indicator [0,1] | Saturday/Sunday Monday – Friday |
| | Sine of week | Sine or Cosine with one full cycle over a week |
| | Cosine of week | |
| | Sine of day | Sine or Cosine with one full cycle over a day |
| | Cosine of day | |
| | Vacations [0, 1] | Indicator if day is a bank holiday |

### 2.4. Correlation Analysis
Ten features were selected as inputs for the ML-algorithms. The correlation between the measured loads and selected features is calculated with the covariance $cov$ (1) and the correlation $corr$ (2) [35].

$$cov(a,b) = \frac{1}{n-1}\sum_{i=1}^{n}((a_i - \overline{a}) \times (b_i - \overline{b})) \quad (1)$$

$$corr(a,b) = \frac{cov(a,b)}{sd(a) \times sd(b)} \quad (2)$$

In equation (1), $a$ and $b$ are the features and $\overline{a}$ and $\overline{b}$ as the means of the features. $sd$ describes the standard deviation of the respective features. To develop an in-depth understanding of the correlations, the data is evaluated once using the full dataset and one time using a monthly separated dataset, to find possible seasonal or calendrical relationships.

### 3. METHODOLOGY
In this section, the load forecasting framework and forecasting metrics are described. A requirement of the forecast algorithm is to be as self-maintaining as possible, with a self-learning behaviour to adopt new patterns and to be independent of seasonal or calendrical effects.

### 3.1. Algorithms

In this study, load-forecasting was computed by means of three different algorithms: Feed-Forward Neural Networks, Long Short-Term Memory and personalized standardized load profiles. To design the machine learning algorithms the framework Keras is used [36].

#### 3.1.1. Feed-Forward Neural Networks

Feed-Forward Neural Networks are basic artificial neural networks that are capable of modelling nonlinear relationships. These neural networks consist of one input layer, one output layer and a variable number of hidden layers. Within these layers are nodes (also called neurons) whose number is variable. The nodes in the input and output layers are typically as high as the number of features used or the number of expected results, respectively. The neurons in one layer are fully connected to those in the next layer by weighted connections ($w_{i,j}$). The input value ($in_j$) of a node is described by the weighted sum of the output values ($a_i$) of the nodes of the previous layer and a bias (b) that can be assigned as optional (2); (for further reading, see Goodfellow et al. [37])

$$in_j = \sum_{i=0}^{n} w_{i,j} a_i + b \qquad (2)$$

To further process the input value, an activation function is used to decide whether to activate a neuron. Activation means that a neuron gives a value to the next layer. The activation function used is the *Rectified Linear Units-Function* (ReLU), which activates neurons when the input value is higher than 0.

The FFNN were designed and trained in this study using the Keras implementation [36].

#### 3.1.2. Long Short-Term Memory

Long Short-Term Memory (LSTM) is an efficient time series modelling architecture that belongs to the *Recurrent Neural Networks* (RNN) methodology. Unlike the above-described neural networks, recurrent neural networks have a feedback connection that allows for the storing of information from recent inputs in the form of activations. Problematic for the training of RNN is the vanishing gradients problem when having long time dependencies. In order to solve this weakness and enhance performance of the RNN, LSTM were introduced by Schmidhuber and Hochreiter in 1997 [38].

The Long Short-Term Memory architecture consists of a memory cell that is connected by an input gate, an output gate and a forget gate. The forget gate ($f_t$) is the first gate in a LSTM unit and controls the information stored within the cell from the last time state ($c_{t-1}$) in accordance with equation (3). The input gate determines which current information is used for the current state (equation (4)), while the output gate controls the amount of information used for the output (equation (5)).

$$f_t = \sigma(W_f \cdot [h_{t-1}, X_t] + b_f) \qquad (3)$$
$$i_t = \sigma(W_i \cdot [h_{t-1}, X_t] + b_i) \qquad (4)$$
$$o_t = \sigma(W_o \cdot [h_{t-1}, X_t] + b_i) \qquad (5)$$

Where a sigmoid activation function is denoted by $\sigma$, the different weights and biases by $W_x$ and $b_x$ of the candidate neuron, being the hidden layer output at time step t-1 and $X$ the input vector at each time step.

The current hidden state of $c_t$ is determined by the following equation:

$$\tilde{c}_t = \tanh(W_c \cdot [h_{t-1}, X_t] + b_c) \qquad (6)$$
$$c_t = f_t * c_{t-1} + i_t * \tilde{c}_t \qquad (7)$$

Where the tanh activation function is denoted by *tanh*, while $W_c$ and $b_c$ denote the weights and bias of the current gate. The output of an LSTM layer is calculated by the following equation:

$$h_t = o_t * \tanh(c_t) \qquad (8)$$

For further information, see Goodfellow et al. [37]. In this study the LSTM implementation from Keras is used.

#### 3.1.3. Standardized Load Profiles

In Germany, standardized load profiles were developed by the "Verband der Elektrizitätswirtschaft e. V." (VDEW) using the load measurements from 1209 different buildings [6]. In contrast to neural networks these profiles are statistical and a low performance approach to load forecasting. In this study it is used as a baseline benchmark to which the other methods are compared to. The significance of it to load forecasting of commercial buildings arises as discussed in section 2.1 and shown in Figure 1, the load patterns within the observed buildings are stable. This was also shown by Edwards et Al. who compared commercial load profiles to residential load profiles [21]. And with that using fixed ruled algorithms are a valuable option to load forecasting in commercial buildings.

In total, there are 11 different standardized load profile sets with a time resolution of 15 minutes. These are divided by the type of customer into: household (1 profile), commercial buildings (7 profiles) and agricultural companies (3 profiles). Every profile set is further subdivided into nine different profile curves, with a differentiation between types of days (weekdays, Saturday and Sunday) and seasons (winter, summer and transition period; Table 3) (Figure 2).

Table 3. Seasonal separation for the standardized load profiles.

| Season | Period |
|---|---|
| **Summer** | 15/05 – 14/09 |
| **Transition** | 21/03 – 14/05 and 15/09 – 31/10 |
| **Winter** | 01/11 – 20/03 |

Public holidays within these periods are treated as Sundays, excepting Christmas and New Year's Day, which are treated as a Saturday if they do not fall on a Sunday. The load profiles are normalized to an annual consumption of 1000 kWh/a and must be adapted by the specific annual consumption of the building under consideration. Forecasting is performed by applying the load profile values according to the season, time and type of day to the forecasting horizon [6].

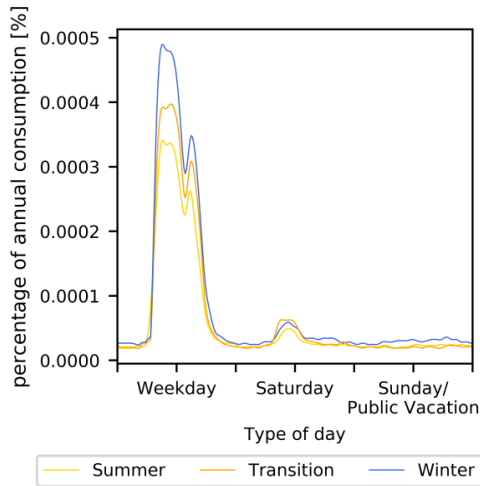

Figure 2. Visualization of the G1 profile as an example of the VDEW standardized load profiles.

As already described several profile sets are available. Because of the characteristics of the load profiles in the observed commercial building the G1 profile is chosen which is for commercial buildings with working hour 8-18 on working days.

#### 3.1.4. Personalized Standardized Load Profiles

An extension of the standardized load profiles are the personalized standardized load profiles. Like the SLP the PSLP is a statistical model, with the advantage of not relying on predefined curves. The load profiles are derived from measured loads form the observed building. In contrast to the SLP these are specific to the building and updatable in regular cycles when new measurements are available.

The preparation procedure for these profiles follow the methodology of the standardized load profiles [6]: Initially, the measured loads are being classified into the same categories as the profiles of the SLP (section 3.1.3).

Afterwards, the profiles are calculated using the mean value for every point of time in the three classes: weekday, Saturday and Sunday/holiday. Exemplary shown in Figure 3 for a weekday profile, where the overlaid grey curves describes the historical measurements of one class and the red line the resulting PSLP. [7]

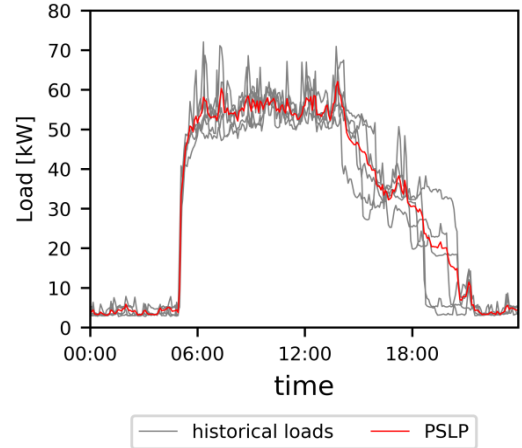

Figure 3. Example of a personalized load profile derived out of measured historical loads in the observed building for weekdays

A general problem of the PSLP approach is that until a full year of load measurements is available, no profiles may be available for every season and day in the first year. To fill this gap, the profiles of the prior season are used in this study. Alike the standardized load profiles forecasting is done by applying the different profiles according to the time of day, specific day and season.

#### 3.2. Metrics

For reasons of comparability classic evaluation metrics are used in this study which all have different meanings and benefits. The RMSE, for instance, has the benefit of penalizing large forecasting errors stronger than the MAE, because the error is squared for every data point on the forecasting horizon. To compare the different methods in detail, four different metrics were used to evaluate the forecast accuracy. These metrics are described in the following equations:

$$MAE = \frac{1}{h} \sum_{t=n+1}^{n+h} |y'_t - y_t| \qquad (9)$$

$$MAPE = \frac{1}{h} \sum_{t=n+1}^{n+h} |\frac{y'_t - y_t}{y'_t}| \qquad (10)$$

$$RMSE = \sqrt{\frac{1}{h} \sum_{t=n+1}^{n+h} (|y'_t - y_t|)^2} \qquad (11)$$

$$MASE = \frac{MAE}{\frac{1}{h-1}\sum_{t=n+1}^{n+h}|y_t - y_{t-7D}|} \quad (12)$$

In the formulas, $y'_t$ is the predicted value and $y_t$ the measured one to a specific time step $n$ and number of time steps in the forecast horizon $h$. Whereas MAE, MAPE and RMSE are used in many studies (e.g., [30] and [31] the MASE (Mean Absolut Scaled Error) is not often used and is further described in [32]. This metric differs from the other methods, in that it is independent of the scale of the data. It compares the MAE reached by the tested method to a naive prediction that is, in this study, the seven-day previously measured power value ($y_{t-7D}$). With a value of the MASE below 1, the method performance used is better than the naïve forecast.

### 3.3. Load-Forecasting Methodology

The concept of the developed load forecast follows the idea of a constant data flow, as illustrated in Figure 1. The forecast algorithm should respond to changed behaviour and provide updated, in-time load forecasts, including the latest available measurements of the building. The methodology is shown in Figure 5 and described in the following sections.

#### 3.3.1. Online Sliding-Window Approach

In an operational environment new data points are constantly collected. With every new data-point, new information is available. Neural networks benefit from more and new information and therefore the forecast accuracy can be improved.

In this study this constant stream is simulated by a sliding window which is moved over the data. The forecast dataset (Figure 4, C) consists of the feature values in the time-period $t$ to the simulation time $t + forecast horizon$ (Figure 4, C). The training dataset always contains the example of the last n-data points within a time period from $t - window\ size$ to the simulation time $t$ (Figure 4, B). All data prior to $t - window\ size$ (Figure 4, A) are not used. After completion the frame is shifted by one simulation step. This concept combines an offline training approach with a constant changing training dataset [39]. In literature this is also referred to as online learning [40].

Similar to this approach is the n-fold-cross-validation where the dataset is split into n partitions. One partition is then selected to be the test dataset whereas the rest of the dataset is used for training. The difference between these two approaches is that by using cross-validation future information is put into the training dataset which would not be available in a real environment. Therefor this would influence the behaviour of the algorithms and it might lead to an increase in accuracy because of foreshadowing of events.

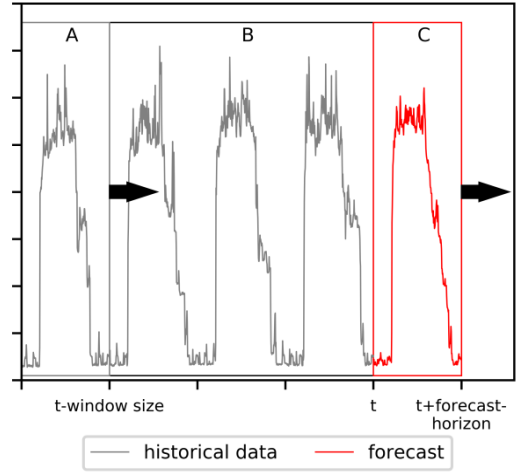

Figure 4. Sliding-window approach simulating a constant stream of data with section A containing unused data in the current simulation step; B as the trainingdata and c the forecast horizon

#### 3.3.2. Data Pre-processing

In the first step of the proposed sliding window methodology within the prediction process (Figure 5. red box no. 1), the dataset must be pre-processed concerning the data quality, time resolution and preparation of an ML problem.

Because quality issues were detected in the data exploration stage (section 2.2) relating to missing data, these must be recovered first. The recovery of missing data points is achieved by linear interpolation.

As already mentioned in the introduction, smart metering systems enable load measurements on building level with decreasing granularity. The incoming data is collected at a time resolution of 1 second and is subsequently transposed to a 5-minute resolution. This is done to lower the amount of load peaks in the data while maintaining a lower forecasting granularity.

For the neural networks, the dataset is normalized by the "MinMaxScaler" provided with the python library Scikit-learn [41], which is based on equation (13). In the formula, $x_{min}$ and $x_{max}$ are the lowest and highest values of the dataset. The data is normalized with $x_{scal}$ to the range between 0 and 1. Descriptive and target features are then normalized separately and the scaler $x_{scal}$ and $y_{scal}$ are cached. To avoid the introduction of new information, the scaler is refitted before each training step of the ML algorithms with the data used for the training of neural networks.

$$x_{scal} = \frac{x - x_{min}}{x_{max} - x_{min}} \quad (13)$$

### 3.3.3. Training and Forecasting

The training process (Figure 5. blue box no. 2) is limited to the length of the input data, or rather the load measurements of the last two months of this study. The compiling of the neural networks is done once at the beginning of the sliding window forecast approach and then fitted to the given data. Afterwards the neural networks always gets refitted but not compiled again. The refitting process of the PSLP was performed every 24 hours at 12 p.m. and, in contrast to the neural networks, implemented with all available data up to the current simulation time step.

Because the ML algorithms use seven days of data as a descriptive feature (data from the previous week, section 2.3), these data is initially available for the PSLP. As the SLP is already fitted to the building using the annual consumption, no further training is conducted.

Forecasting is performed at every time step as an intra-day update.

### 3.3.4. Evaluation

For the evaluation (Figure 5, grey box no. 3), the predictions are first denormalized by using the cached scaler from the pre-processing stage (section 3.3.2). Afterwards, the evaluation metrics described in 3.2 are used and cached across the entire simulation for every time step and saved at the end.

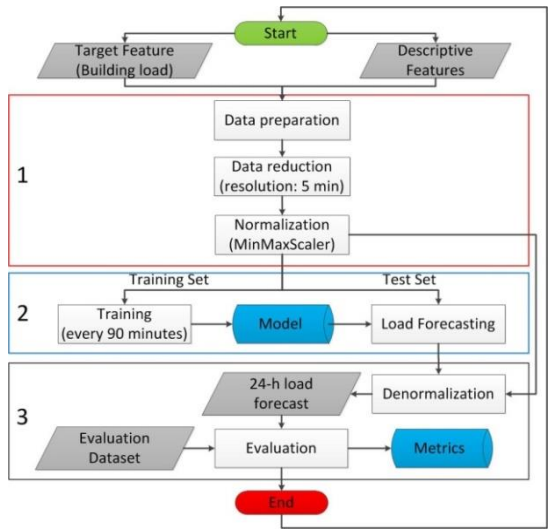

Figure 5. Simplified flow chart of the automated rolling forecast methodology.

### 3.4. Algorithm Implementation and Optimization Procedure

Neural Networks have many parameters that must be optimized in order to gain more accurate forecasts; a complete optimization will not be conducted in this study. The neural networks were designed using Keras [34] in Python 3.6 and optimized according to the number of layers and of neurons per layer (Table 4). To prevent the training from being completed at a local minimum, but also to limit the process time, a patience factor is set at 50. The patience factor stops the iteration after the chosen number of 50 iterations, when no better weight combinations were found and the best weights are restored.

Table 4. Optimization parameters for the ML algorithm network architectures.

| Network architectures | |
|---|---|
| Number of Layers | 1 – 8 |
| Number of Neurons | 8, 16, 32, 64, 128 |
| Loss function | MAE |
| Optimizer | ADAM |
| Activation function | ReLU |
| Epochs | 2000 |
| Patience | 50 |

The testing procedure is equivalent to the method described above but as mentioned in section 3.3, has a shortened dataset containing the data of nearly two months (12/02/2019-04/04/2019).

The simulations were conducted on a server with 2 Intel Xeon E5-2630v4 with 2.20 GHz that have 10 Cores each and can handle 20 threads by using Hyper-Threading and 256 GB of ram.

### 4. CASE STUDY: Integration Of BEVs

In this section, details of the case study are described, demonstrating the integration of battery-electric vehicles (BEVs) into existing commercial buildings as an example for the use of load-forecasting-based energy management.

### 4.1. Case Study Description

The integration of BEVs into existing buildings is a challenging task. The overlap of the typically higher loads in a commercial building during working hours and the newly introduced high charging loads of the connected electric vehicles will require charging strategies to avoid overloading events. In this case study, 22 kW electric vehicle charging stations are integrated into the MD building (Figure 6) with its own smart meter. The charging stations are available as semi-public usable chargers that can be used by workers in the commercial building during the weeks and with no restrictions on the weekend. The goal of the system is to charge the electric vehicles as quickly as possible without compromising the buildings' base loads within the grid connection limits. To achieve these objectives, charging is scheduled on the basis of the developed load forecasts. For the simulation of the case study, six different scenarios are defined, with two, five and 10 charging stations integrated and using them with or without scheduled charging.

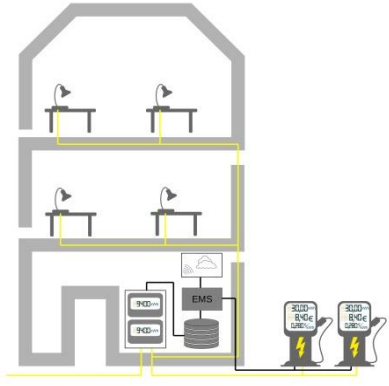

Figure 6. Scheme of the integrated charging infrastructure in the commercial building.

### 4.2. Charging Strategies
Several charging strategies are available for charging BEVs. This study focuses on uncontrolled charging and grid-oriented charging. In the uncontrolled charging strategy, the electric vehicles are always charged with the maximum possible power. In contrast, the grid-oriented charging strategy takes the load within the building grid into account so as to not exceed the grid limitations. This is achieved by lowering the charging load of the connected charging vehicles in light of the load forecast for the next 24 hours.

### 4.3. Simulation of Battery-Electric Vehicle Charging
Simulating the charging process of BEVs is simplified in this study by assuming a constant charging power that will be reduced if necessary. In this case-study also no seasonal effects on the batteries are taken into account and sorely the power is observed. The charging schedule for the grid-oriented strategy is calculated using the load forecast provided by the energy management system. In a first step, the free capacity is calculated as the difference between the maximum grid connection capacity and the forecasted loads. The maximum grid connection capacity is set by the assumption that the maximum load measured equals 80%. For the MD, it is rounded to 110 kW. The free capacity is then split for every time step in the prediction between all vehicles according to their state of charge (SoC), the time connected to the charging station and the maximum charging load of the vehicle. With every prediction, the charging schedule is also updated.

### 4.4. Simulation of Arriving and Departing Vehicles
For the simulation of arriving and departing times of BEVs at commercial buildings, a synthetic dataset is generated.

10 different profiles were randomly designed with the following parameters: different driven distances to work, different driving behaviour on weekends and different arrival and departure times at the commercial building.

To randomize the behaviour of the drivers, an offset is randomly applied to each category "every day." While the first person arrives when the loads increase in the building, the arrival time is set to the exact time when the observed load profile rises (Figure 1). The same procedure is applied to the last person who leaves the building, which is set to the time when the load decreases. The type of BEV is randomly assigned, with battery capacities between 18.7 kWh and 100 kWh and a charging power of 11 kW or 22 kW.

On the weekends, private citizens charge their cars at the stations. The chance of one person arriving at the building on the weekend between 8 a.m. and 10 p.m. is assumed to be 5%. In contrast to the profiles of the employees, the state of charge (SoC) of the BEV is randomly calculated in the range of 5% to 20%.

The result of the presented process is shown as a heat map in Figure 7.

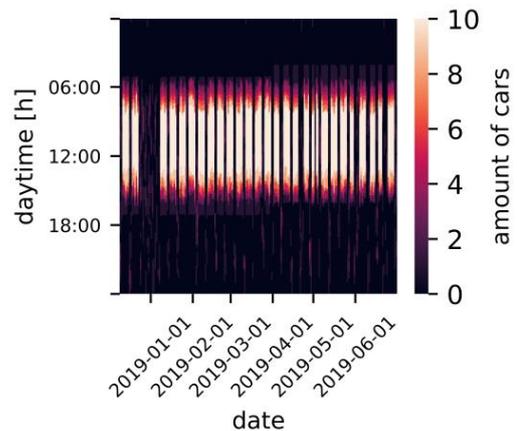

Figure 7. Heat map of the synthetically-derived arriving and departure dataset of vehicles in the building.

## 5. RESULTS: Load Forecasting
In this section, the results concerning the load forecasts for commercial buildings are described and discussed. The results are compared to a fourth method with the standardized load profiles as benchmarks.

### 5.1. Feature Analysis
The correlation between descriptive features and measured load can have seasonal dependencies. Figure 8 illustrates the correlation between descriptive features and the measured loads split in the month and for the entire dataset.

The correlation analysis demonstrates that the correlation between load measurements and descriptive features partially depends on seasonal effects. The features with the highest correlation directly derived from the dataset (Measurement of the week before, Measurement of the day before). The lower correlation of the *Measurement of the day before* is partly due to the fact that, for example, on a Saturday the load values of Friday are used. Lower correlations are calculated for the independent sine and cosine features and the day and weekend indicators. In contrast to Cai et al. [33], the analysis shows fewer correlations between the temperature and load measurements with stronger monthly fluctuations. Because the models derived by neural networks rely on the training data, fluctuations in the correlations can have an impact on the predictions. Therefore, updating neural network weights by refitting the network to new data is integrated into the workflow (section 3.3, Figure 5). Also described in 2.4, the Pearson coefficient is quantifying linear correlations. Higher degrees of correlations are therefore not investigated.

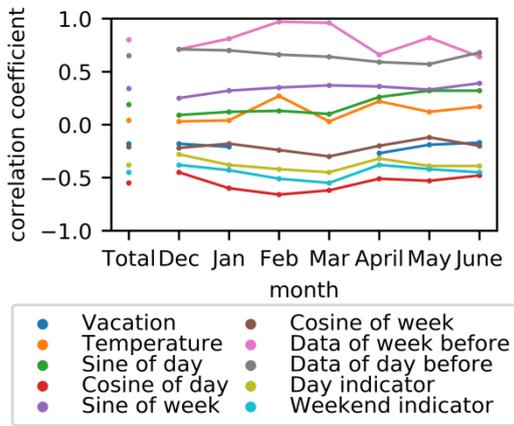

Figure 8. Pearson Correlation between the load measurements and selected features.

### 5.2. Initial Prediction Error

In contrast to the SLP, the PSLP and ML algorithms depend on the availability of load measurements from the building. As the ML algorithms rely on the feature, "Data of the last week," these data are initially present for the PSLP. This leads to the PSLP being able to predict loads from the beginning of the simulation onwards in this study. The ML algorithms, in contrast, have an initial prediction error, as demonstrated in Figure 9. The characteristics of this are two periods of high forecasting errors within the first week of the simulation. The first peak is at the beginning of the simulation, as long as less data are available to derive a model (marker 1 and 2, lower graph, Figure 9). The second peak is reached when the first weekend occurs. Therefore, it is predicted by the network as a weekday (marker 3, Figure 9). This behaviour must be taken into account when deploying a load forecast based load management system. This is due to the high forecasting error of an untrained neural network on less data. The length of this event can vary but it is approximately about one combination of two working days and a weekend. With this procedure is a one-time event in the simulation, the assessment of activity in the following sections is adjusted evaluating the performance after the initial period.

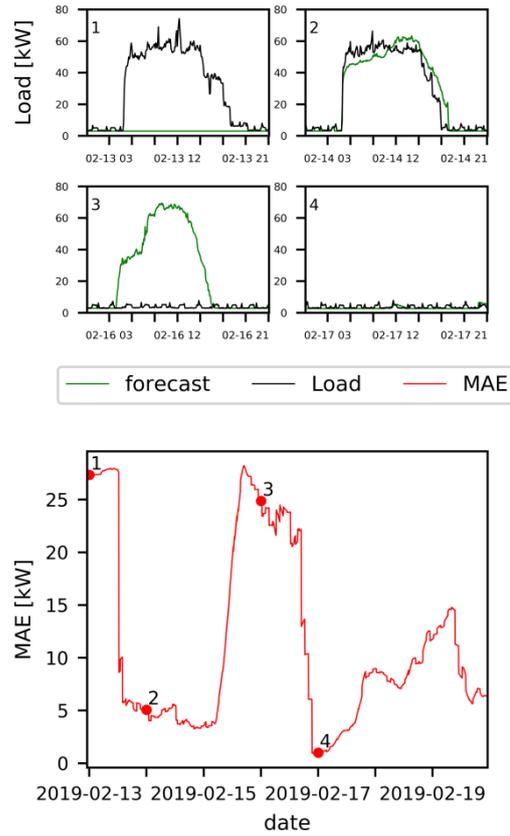

Figure 9. MAE in the initial phase of the simulation of an LSTM with three layers and 16 neurons.

### 5.3. Neural Network Architecture Optimization

Choosing a suitable architecture and parameters is an important step towards improving the forecast accuracy. An excerpt of the simulation results for different neural network architectures is shown in Table 5 (full table in Appendix A.2). The metrics (MASE, MAPE and RMSE) are listed as averaged over the test period (02/03/2019-03/04/2019).

Table 5. Excerpt of the optimization results. Forecast accuracy of different neural network architectures. (averaged values of evaluation metrics)

|  | Configuration | MAE [W] | MASE | RMSE [W] |
|---|---|---|---|---|
| FFNN | 1 Layer, 8 Neuron | 3583 | 0.97 | 6198 |
| FFNN | 4 Layer, 8 Neuron | 3508 | 0.92 | 6492 |
| LSTM | 2 Layer, 8 Neuron | 3497 | 0.91 | 6197 |
| LSTM | 7 Layer, 8 Neuron | 3426 | 0.88 | 6333 |

The results demonstrate that both neural networks tend to perform worse with more neurons than the tested minimum number of eight neurons. Even an increase of hidden layers beyond four did not lead to a better performance, aside from the seven-layer and eight-neuron LSTM architecture. The MASE revealed that most neural network architectures with more layers and neurons are mostly worse than a persistency prediction. This also shows that although having stable load patterns, more sophisticated approaches like neural networks can improve forecast accuracy.

In total, there are several feed-forward neural network configurations with comparable results. The best results could be achieved with four layers and eight neurons regarding the MASE and MAE, but worse for the RMSE compared to a network with one hidden layer and eight neurons. For this study, the more complex, with four hidden layers and eight neurons, was chosen because of its lower MAE and the much lower MASE compared to the one with 1 layer and 8 neurons.

For the LSTM the 7 Layer and 8 Neuron architecture was chosen as like for the FFNN this architecture has the lowest MAE and MASE whereas the RMSE is slightly worse.

Comparing both of the selected neural network architectures, the LSTM outperforms the FFNN by having higher forecast accuracies on all scales. In the next section the neural network architectures are used on longer dataset containing vacations and it is assessed how they perform against the SLP and PSLP as their benchmarks.

**5.4. Comparison of the Load-Forecasting Methods**

The previously optimized ML algorithms are used to evaluate the real performance on the full main dataset. This includes public vacations and different seasons. The averaged results are shown in Table 6.

Table 6. Comparison of the algorithm performance on the whole MD (averaged values of evaluation metrics).

|      | MAE [kW] | RMSE [kW] | MAPE [%] |
|------|----------|-----------|----------|
| FFNN | 4.11     | 8.35      | 50.34    |
| LSTM | 4.47     | 9.29      | 51.42    |
| PSLP | 3.99     | 9.09      | 53.22    |
| SLP  | 9.26     | 16.65     | 69.4     |

In comparison to the SLP, the data-based algorithms perform significantly better than the SLP. The FFNN perform best on the MAPE and RMSE evaluation metrics, but is outperformed by the PSLP on the MAE. Within the optimization stage, the LSTM performs worse than the FFNN and is also outperformed by the PSLP. The averaged MAEs are within a range of around 4.7% to 5.3% of the peak load for the data-driven algorithms, which is compared to the 11% for the SLP measured a significant improvement.

Figure 10 presents the boxplots of the forecasting errors to have a deeper view on the error distribution. A boxplot consists of a box, or the so-called interquartile range, the whiskers (upper and lower lines), the median (orange line) and usually outliers, which are not shown in the figure. The interquartile range contains 50% of all the values. Value errors higher or lower the interquartile range are described by the whiskers. All other data points are considered outliers. The boxplots show that the SLP has the largest error distributions in all three metrics with a large interquartile range and a high distance between the box and the upper whisker.

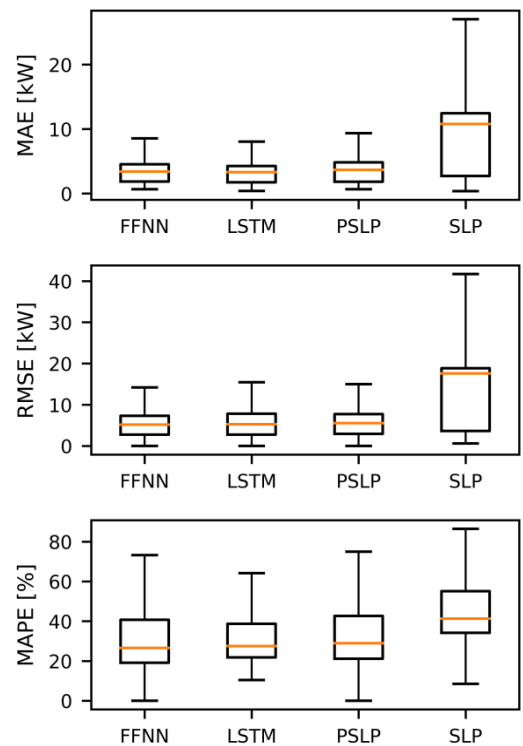

Figure 10. Comparison of the error metrics (MAE, RMSE and MAPE) of the ML and standardized load profile approaches. The boxes show the interquartile range which contains 50% of the values whereas values higher or lower and described by the whiskers. The orange line indicates the median.

Significant for all of the three methods is the high amount of outliers present beyond the upper whiskers (maxing out at: MAE 34 kW, RMSE: 205 kW, MAPE: 870 %; on public holidays). The ML approaches slightly outperform the PSLP on the MAE and RMSE scale by having a smaller interquartile range, a comparable or lower upper whisker and lower median (orange lines). This shows that both can predict the future loads more consistent and more accurate as the PSLP. In contrast to the MAE, FFNN and PSLP performing similar on the MAPE metric, the LSTM has a higher lower whisker and a lower interquartile range while maintaining a lower upper whisker.

As mentioned in section 5.2, there is an initial prediction error of the ML algorithms in the first week (Appendix A.1). In total, the differentiation between predictions on weekdays and weekends is clearly outlined by the height of the prediction error. On weekends, the MAE is typically lower than the MAE on weekdays due to the lower and less fluctuating loads.

To further evaluate the behaviour, a closer look is taken into the prediction of a weekday (Figure 11). In contrast to the increase of the load demand in the morning (4 a.m.), which has a higher degree of regularity, smaller load peaks and the decrease in the load cannot be predicted so accurately (5 p.m.). It is also shown that compared to the PSLP (orange), both neural networks (FFNN: blue; LSTM: green) tended to predict constant loads during working hours.

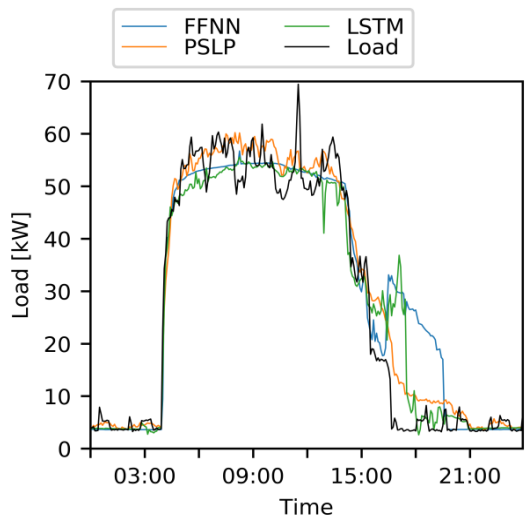

Figure 11. Example of the predictions of the data-driven algorithms for load-forecasting (date: 28/05/19).

### 5.5. Adaptability of the ML-Algorithms / Public Holidays

Adapting new patterns must be performed automatically by the load-forecasting algorithm in order to provide accurate forecasts and to be as self-maintaining as possible. In this study is evaluated by using the Christmas holidays as an example of an abrupt change in user load demand behaviour in the building.

Regarding Figure 12, the ML-algorithms automatically adapt the new behaviour. The adaptation process for both algorithms has three steps. The first step is characterized by a high prediction error (Figure 12, marker 1). The adaptation stage (step 2), which can lead to an abruptly decrease like in case of the FFNN or has an increased MAE (LSTM) again. In the last stage the adaptation is completed (Figure 12, marker 3) until the next public holiday on the 1$^{st}$ of January.

In contrast to that, the PSLP and SLP follow fixed rules, described in sections 3.1.3 and 3.1.4, and do not adopt the new patterns automatically. Both have a high forecasting error, because the prediction on December 27$^{th}$ was categorized as a working day (Figure 12, marker 2). This is a problem when using methods with fixed rules because they must be monitored manually or new and personalized rules have to be implemented so that they can react to unforeseen changes.

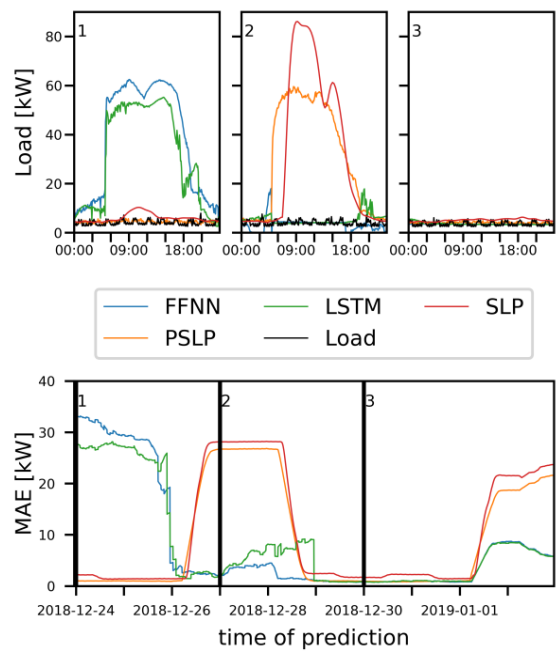

Figure 12. Prediction errors while adapting to new behaviour.

In contrast to the neural networks, events like public vacations can be accurately predicted by the PSLP and SLP if they are specified. The ML algorithms can predict vacations as well, but as they are a data-driven approach, they require examples in the dataset (Figure 13). As the training dataset does not include vacations, the prediction error arises when the first vacation date occurs (Figure 13, marker 1). When the new situation arises, the FFNN predict even negative values, although no negative values are available in the dataset. Unlike the LSTM, the prediction on the second public vacation some weeks later, the FFNN predicts the load demand more accurately, whereas the LSTM still predicts high loads (Figure 13, marker 2).

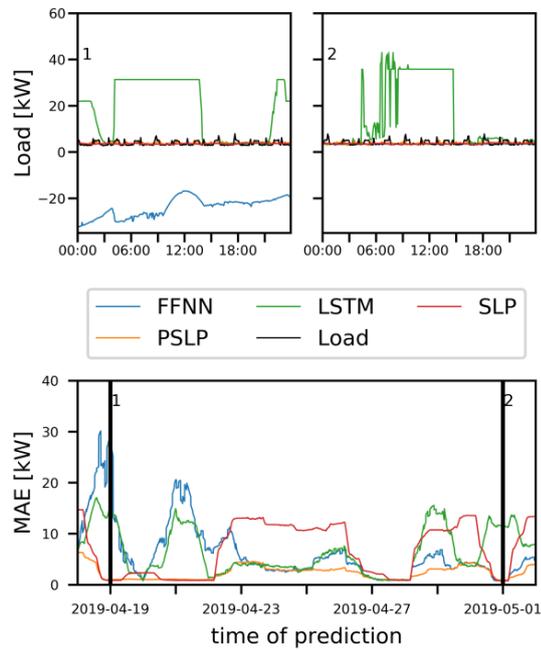

Figure 13. Forecast errors when predicting public holidays with and without an example in the training dataset

## 5.6. Validation of the results

Optimizing neural networks for new building is time consuming and requires long time-series of measured loads from the new buildings. Therefore in this section non pre-trained neural networks are used on the validation dataset (VD) which have the same architecture like the once used for the MD in the previous sections. A comparison of the results is illustrated in Figure 14.

The MAPE reveals that the PSLP (orange) performs slightly better on the MD than on the VD while the ML algorithms (LSTM: green; FFNN: blue) performing better on the VD. The same is shown by the MASE in case of changing load patterns, described in section 2.1, the persistency forecast accuracy is decreased and therefor the MASE also decreased for all algorithms.

As is shown in sections 5.2 and 5.4, the initial prediction error also appears, when predicting on the basis of the VD dataset (Figure 15). It was pointed out in 2.1 that the loads on the weekends differ significantly between the MD and VD. This can also be observed in the MAE profile, with the weekends not as visible in the VD as the MD. It is demonstrated that the behaviour of the algorithms used is comparable in both datasets, with a slight exception in the PSLP. This algorithm sometimes has a significantly higher or lower MAE compared to the ML-algorithms on the VD, while the MAE of all algorithms is more equal on the MD.

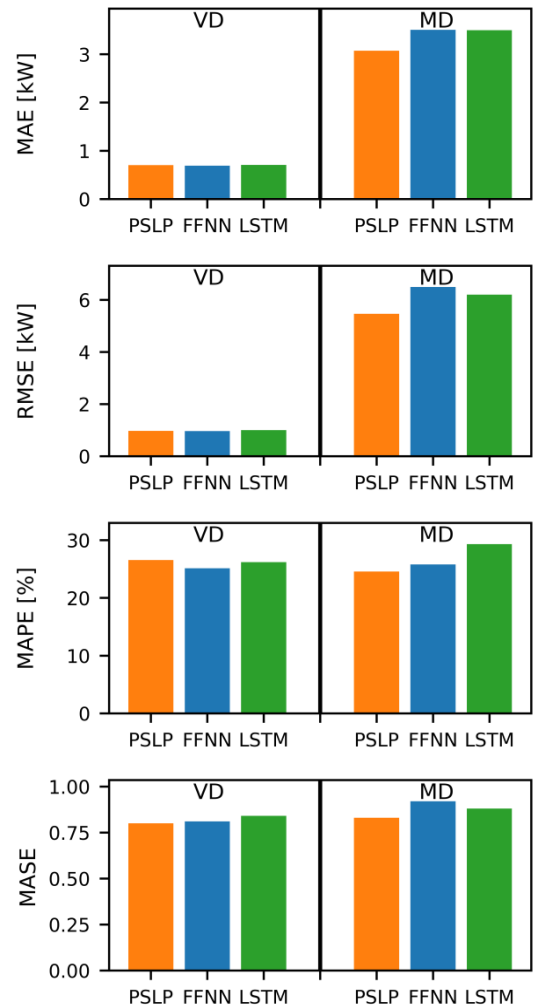

Figure 14. Comparison of evaluation metrics for validation research on MD and VD

Due to the differences in the results regarding the MASE and MAPE it is better to optimize the neural network architecture once again for the VD. Therefore using ML-approaches for load management applications is more complex than using simple PSLP. But the potentials of having even higher forecasting accuracies with more sophisticated algorithms are important to accurately manage loads in a self-maintaining environment. The ability to use the same neural network architecture for different datasets/buildings can be a key factor for fast deployment, as well as the use of pre-trained algorithms. This would lower the complexity of the problem deploying a load forecast based energy management system, so that it is usable without having measurements over a long time period and to avoid of individual optimization work.

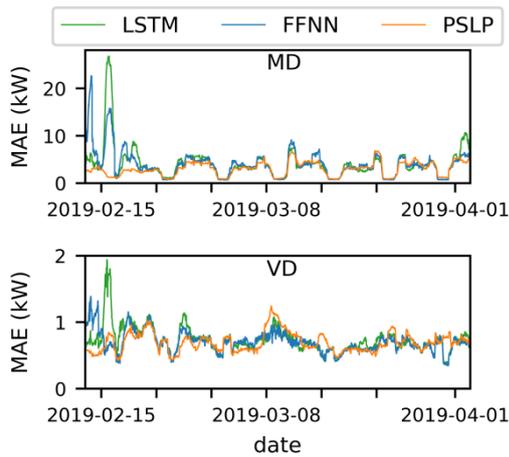

Figure 15. Comparison of the MAE profile of the predictions made on the MD and VD.

## 6. RESULTS: Case Study

The scenarios described in section 4 were simulated using the shortened MD. The FFNN is used as a forecast algorithm.

According to Table 7, the integration of two charging stations did not show any problem, as no overload was registered and the system did not influence the charging procedure. This changes for the scenario with 5 charging stations, as the average charging duration increased by about four minutes. The calculated mean and maximum overload for the controlled charging reflect how high the overload of the grid connection would be if the energy management system were to charge as scheduled. Because of the scheduled charging, both significantly decreased. This is also shown in the simulated scenario with 10 charging stations. As a result of the larger amount of vehicles and the limited available capacity for all vehicles, the averaged charging duration increases by over 30 minutes.

To further evaluate the issue, Figure 16 points out that if the load prediction over-estimates the loads, the scheduled charging loads will be significantly lower and therefore will not charge at the limit and so will not exceed it. For underestimation, which is shown in Table 7 by the registered overloads, the opposite occurs. Also the initial prediction error which was researched for the neural networks is also applicable to the PSLP as with no data available no predictions can be done and the system cannot manage the loads.

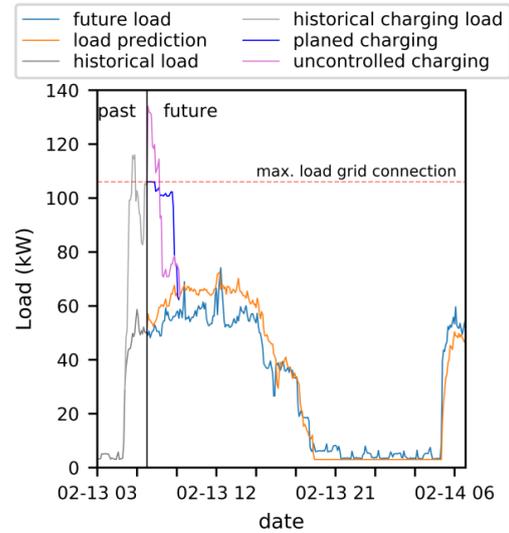

Figure 16. Comparison between controlled and uncontrolled charging in an energy management system with load forecasting

Another issue arises regarding the change of behaviour within the building when for example tenants change. In section 5.5 and 5.6, with regards to the load pattern characteristics described in 2.1, showed that the ML-approaches are capable of adopting new patterns. In contrast to that statistical approaches need new rules to be able to (see section 5.5). In both cases the changes were not significant. A more significant change could be in a multi-customer commercial building, when one tenant with high loads or several tenants move out. This can lead to significant changes and can result in high forecasting errors which will be reduced automatically by the ML-approaches. For the PSLP a reset of the load profiles might be needed to adapt major changes in the load behaviour. Anyway, both methods lead to potentially unavoidable high forecasting errors what affects the scheduling of dynamic loads. High forecasting errors on public holidays are unavoidable too, but as commercial buildings tend to have low loads on these days, the problem slightly distinguishes compared to high forecast errors on weekdays. The initial prediction error must always take into account if ML-techniques were chosen.

Table 7. Results of the simulations of the integration of electric vehicles by using load forecasts to control charging

| Number of charging stations | 2 | | 5 | | 10 | |
|---|---|---|---|---|---|---|
| Scenario | Controlled | Uncontrolled | Controlled | Uncontrolled | Controlled | Uncontrolled |
| Average energy charged [kWh] | 24.75 | 24.75 | 23.41 | 23.41 | 23.84 | 23.84 |
| Average charging duration [hours] | 01:26:02 | 01:26:02 | 01:36:09 | 01:32:37 | 02:10:17 | 01:32:22 |
| Registered overloads | 0 | 0 | 7 | 90 | 605 | 826 |
| Maximum Overload [kW] | 0 | 0 | 9.7 | 24.9 | 20.24 | 65.15 |
| Mean overload [kW] | 0 | 0 | 3.38 | 6.86 | 3.91 | 23.89 |

All of that leads to the fact, that load forecasting approaches can be used to estimate charging schedules for BEV. But the aforementioned characteristics show also, that the different used approaches have different benefits and shortcomings for load management applications.

Additionally, a load forecast based energy system can provide the electric vehicle's owner with information if charging the vehicle is possible and give the owner of the charging station new opportunities for private-public charging stations without compromising the base load of its building and the implementation of new tariff systems.

## 7. CONCLUSION AND OUTLOOK

In this study, the performance and accuracy of traditional load forecast methods as the SLP and PSLP was compared to basic neural networks: LSTM and FFNN. As a load forecasting framework, a rolling load forecasting methodology was proposed to simulate the integration of a load forecasting algorithm in an energy management system. The evaluation of the behaviour from different forecast algorithms was also investigated in this work. The results of the neural network forecast algorithms performed more accurate in the optimization stage using an idealized dataset (without holidays) compared to the not idealized dataset (section 5.5). In total, the FFNN and LSTM demonstrated the ability to automatically adopt new behaviour patterns. Also important is the initial prediction error (section 5.2), which occurred at the very beginning of the simulation and featured high prediction errors.

In contrast, the PSLP and SLP cannot dynamically respond to changing events due to fixed rules on, e.g., the time between Christmas and New Year's Day (section 5.5). With more rules regarding these exceptions and the results, the PSLP is a reasonable alternative to ML algorithms in forecasting loads for single commercial buildings. The PSLP provides comparable forecast accuracy (section 5.4), but also a lower training process duration (Table 6).

As demonstrated, the same algorithms can also predict loads in another building, but the forecast accuracy is not the same (Section 5.6).

As a use case, the usage of load-forecasting in the energy management system of an existing commercial building to integrate BEV charging infrastructure was designed and simulated. Especially for the use-case the granularity of the data was raised to 5 minutes to ensure more accurately scheduling and also the ability to predict load peaks. The results indicate that this can ensure the integration of more charging stations while having a small margin between building load and grid connection limit, without extending the existing infrastructure.

It is still important that the energy management system always measures and limits the loads in time, as the load forecast can provide a raw estimation of how high the loads will be. In summary, load-forecasting can be used to shift loads or at least provide information when these consumers can be used to not compromise the base loads of the building. Also in combination with a PV power prediction the self-consumption can be increased and therefor the loads of grid connection get relieved.

Load-forecasting at the building level does not ultimately require ML algorithms to ensure accurate predictions. Nevertheless, the full potential of ML algorithms must be further evaluated and more sophisticated neural network architectures have to be applied to the problem of single building level forecasting. As discussed in the validation optimizing neural networks to new buildings could lead to better forecasts but this has to be evaluated in further studies. By Combining statistical and ML-approaches a reliable load management system could be introduced to the problem. As an example the PSLP could support the ML-Algorithm by bridging the gap between the first measurements until accurate predictions are available and during public vacations. It is also possible to use the PSLP as a feature for training of neural networks as these have proven to be valuable to forecast loads in a commercial building. Also the PSLP could bridge times between training of neural networks if it is not finished in time or problems arise like false forecasts or other issues. In total the energy management system must still decide and regulate loads in-time in order to ensure maximum charging power to every point in time and prevent overloads.

Related to the framework, the algorithms must be tested in an operational energy management system environment to further evaluate the problems arising by using ML algorithms and PSLP for load-forecasting. This also applies to the use case of integrating electric vehicles into an existing building, as many assumptions had to be made.

In future research the focus should also be taken onto the change of the needed neural network architecture. As an option to implement hyperparameter optimization in an operational environment cloud services could be a key-factor. Or even going further the whole training can be done in the cloud with the trained model used locally for forecasting.

As is demonstrated by the use case, load forecasts can support energy management systems. The forecast-based operational strategies of flexible consumers and battery storage capacities must be researched. Concerning the integration of the three main sectors, the integration of electrical space heaters or heat pumps is also made possible by combining the forecasts of heat usage and load-forecasting.

## 8. ACKNOWLEDGEMENTS

The authors acknowledge the financial support of the "Federal Ministry for Economic Affairs and Energy" of the "Federal Republic of Germany" for the project "EG2050: EMGIMO: Neue Energieversorgungskonzepte für Mehr-Mieter-Gewerbeimmobilien" (03EGB0004G and 03EGB0004A). For more details, visit www.emgimo.eu. The presented study was conducted as part of this project.

# Appendix

## A.1 Total results of the neural network optimization stage

Table 8 Averages of MASE, MAE, RMSE for all neural network architectures tested for optimization

| | | Layer/Neuron | 1 | 2 | 3 | 4 | 5 | 6 | 7 | 8 |
|---|---|---|---|---|---|---|---|---|---|---|
| MASE [-] | FFNN | 8 | 0.97 | 0.96 | 0.94 | 0.92 | 0.96 | 0.97 | 0.96 | 1 |
| | | 16 | 1.04 | 1.1 | 1.17 | 1.03 | 1.11 | 1.04 | 0.98 | 0.99 |
| | | 32 | 1.12 | 1.12 | 1.09 | 1.29 | 1.03 | 1.13 | 1.02 | 1.07 |
| | | 64 | 1.14 | 1.16 | 1.08 | 1.12 | 1.2 | 1.09 | 1.04 | 1.05 |
| | | 128 | 1.2 | 1.25 | 1.43 | 1.32 | 1.17 | 1.24 | 1.19 | 1.09 |
| | LSTM | 8 | 0.91 | 0.91 | 0.93 | 0.94 | 4.9 | 4.9 | 0.88 | 1 |
| | | 16 | 1.04 | 1.08 | 1.14 | 1.15 | 1.02 | 1 | 4.9 | 4.9 |
| | | 32 | 1.03 | 1.02 | 1.16 | 1.32 | 1.15 | 1.16 | 1.1 | 1.05 |
| | | 64 | 1.06 | 1.03 | 1.14 | 1.42 | 1.14 | 1.13 | 1.19 | 4.84 |
| | | 128 | 1.05 | 1.09 | 1.3 | 4.9 | 1.18 | 1.24 | 4.9 | 1.16 |
| MAE [W] | FFNN | 8 | 3583 | 3603 | 3534 | ==3508== | 3702 | 3731 | 3655 | 3786 |
| | | 16 | 3815 | 4192 | 4203 | 3948 | 4100 | 3855 | 3759 | 3715 |
| | | 32 | 4068 | 4087 | 4103 | 4427 | 3797 | 4152 | 3763 | 3937 |
| | | 64 | 4159 | 4173 | 3999 | 3950 | 4309 | 4044 | 3893 | 3913 |
| | | 128 | 4345 | 4456 | 4682 | 4657 | 4282 | 4269 | 4259 | 3973 |
| | LSTM | 8 | 3540 | 3497 | 3525 | 3661 | 20250 | 20250 | ==3426== | 3875 |
| | | 16 | 3817 | 3960 | 4366 | 4230 | 3798 | 3816 | 20250 | 20250 |
| | | 32 | 3890 | 3780 | 4275 | 4676 | 4190 | 4300 | 4025 | 3937 |
| | | 64 | 4099 | 3906 | 4285 | 4853 | 4076 | 4251 | 4199 | 19151 |
| | | 128 | 3973 | 4059 | 4494 | 20250 | 4227 | 4425 | 20250 | 4267 |
| RMSE [W] | FFNN | 8 | ==6198== | 6534 | 6372 | 6492 | 6563 | 6746 | 7017 | 7100 |
| | | 16 | 6571 | 7932 | 7634 | 7145 | 7331 | 7037 | 6758 | 6748 |
| | | 32 | 7059 | 7150 | 7434 | 8469 | 6734 | 7628 | 6884 | 7180 |
| | | 64 | 7153 | 7192 | 6906 | 7058 | 7939 | 7265 | 6941 | 7016 |
| | | 128 | 7624 | 7647 | 8740 | 8477 | 7672 | 7960 | 7895 | 7341 |
| | LSTM | 8 | 6327 | ==6197== | 6418 | 6955 | 30349 | 30349 | 6333 | 7389 |
| | | 16 | 6772 | 6961 | 8505 | 7656 | 6729 | 6848 | 30349 | 30349 |
| | | 32 | 6902 | 6586 | 8845 | 9874 | 7768 | 7980 | 9372 | 7240 |
| | | 64 | 7760 | 6894 | 7607 | 10997 | 7486 | 8404 | 7829 | 28321 |
| | | 128 | 6918 | 7012 | 7928 | 30349 | 7953 | 12828 | 30349 | 139320 |
| Average training Time [s] | FFNN | 8 | 50 | 35 | 42 | 61 | 60 | 50 | 58 | 35 |
| | | 16 | 40 | 47 | 48 | 31 | 42 | 58 | 39 | 34 |
| | | 32 | 44 | 42 | 30 | 51 | 44 | 49 | 45 | 37 |
| | | 64 | 40 | 46 | 45 | 31 | 31 | 68 | 64 | 38 |
| | | 128 | 46 | 42 | 31 | 73 | 29 | 56 | 43 | 36 |
| | LSTM | 8 | 74 | 127 | 110 | 156 | 246 | 257 | 232 | 175 |
| | | 16 | 75 | 70 | 108 | 146 | 129 | 158 | 321 | 229 |
| | | 32 | 65 | 70 | 112 | 137 | 124 | 163 | 292 | 283 |
| | | 64 | 77 | 66 | 85 | 161 | 93 | 89 | 198 | 274 |
| | | 128 | 58 | 79 | 84 | 162 | 126 | 226 | 134 | 124 |

**A.2 Comparison of the MAE of the predictions of the algorithms used at the beginning of simulation**

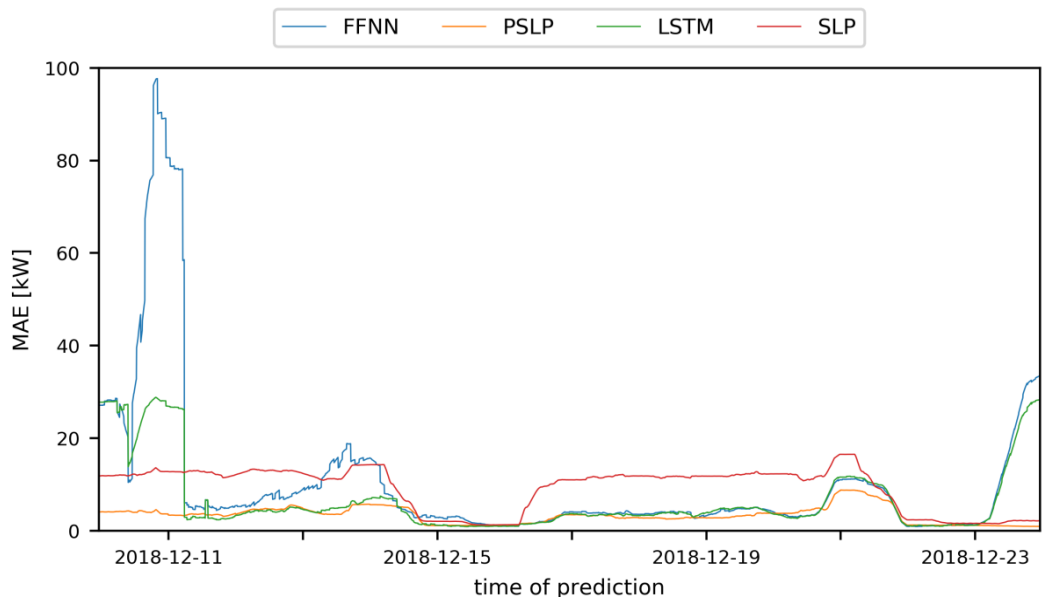